\documentclass[letterpaper, 12pt]{article}
\usepackage{latexsym}
\usepackage{amsmath}
\usepackage{amssymb}
\usepackage[latin1]{inputenc}
\usepackage{graphicx}
\newcommand{\be}{\begin{equation}}
\newcommand{\ee}{\end{equation}}
\newcommand{\bea}{\begin{eqnarray}}
\newcommand{\eea}{\end{eqnarray}}

\newcommand{\bbibitem}[1]{\bibitem{#1}\marginpar{#1}}

\def\Label#1{\label{#1}%
  \smash{\hbox to0pt{\raise1ex\hbox{\tiny[#1]}\hss}}}
\def\noLabels{\let\Label=\label}
\def\nobbibitem{\let\bbibitem=\bibitem}

\title{{\bf On the existence of supergravity duals to D1--D5 CFT states}}

\author{Klaus
Larjo$^\sharp$\footnote{klarjo@physics.upenn.edu}
\\[1mm]
\small \sl $^\sharp$\; David Rittenhouse Laboratories, University of
Pennsylvania,
\\[-1.5mm]
\small \sl Philadelphia, PA 19104, USA \\
}
\date{}

\begin{document}

\noLabels \setlength{\baselineskip}{16pt} \nobbibitem
\begin{titlepage}
%
%\rightline{\small hep-th/yymmnnn}
%\rightline{\small UPR-T-nnnn}
%\rightline{\small UCSB-....}
%\rightline{\small UCB-PTH-07/nn}
%\rightline{\small LBNL-nnnn}
\maketitle
%\rightline{\s
%\begin{picture}(0,0)(0,0)
%\put(350, 420){UPR-T-nnnn}
%\put(350,403){UCB-PTH-07/nn, LBNL-nnnn}
%\put(350,386){hep-th/yymmnnn}
%\end{picture}
%\vspace{-36pt}

\begin{abstract}
We define a metric operator in the $\frac{1}{2}$-BPS sector of the D1-D5 CFT, the eigenstates of which have a good semi-classical
supergravity dual; the non-eigenstates cannot be mapped to semi-classical gravity duals. We also analyse how the data defining a
CFT state manifests itself in the gravity side, and show that it is arranged into a set of multipoles. Interestingly, we find
that quantum mechanical interference in the CFT can have observable manifestations in the semi-classical gravity dual. We also
point out that the multipoles associated to the normal statistical ensemble fluctuate wildly, indicating that the mixed thermal
state should not be associated to a semi-classical geometry.
\end{abstract}
\thispagestyle{empty}
\setcounter{page}{0}
\end{titlepage}

%\newpage

\section{Introduction}
\label{intro}

Recently there has been considerable progress in using the AdS/CFT correspondence to understand quantum gravity, especially in
the form of explicit mappings from certain CFT's to their dual semi-classical geometries. The first such system was the set of
LLM geometries: an explicit map from states of $\frac{1}{2}$-BPS sector of $\mathcal{N}=4$ SU($N$) SYM to their dual supergravity
solutions \cite{llm}. This map was further developed and analysed among others in \cite{vijayjoan,us,usbig, mandal,pettorino,silva1}.
Progress in extending this mapping to the $\frac{1}{4}$- and $\frac{1}{8}$-BPS sectors has been made in
\cite{linnew,donos1,donos2,milanesi,kim}. Another such map was proposed between the $\frac{1}{2}$-BPS sectors of the D1-D5 black hole and
its dual field theory in \cite{jan}, and also in \cite{skenderis1,skenderis2,silva2}; while similar mappings were introduced and
analysed for the set of Lin--Maldacena geometries in \cite{lin,mark}. All of these mappings lend support to the proposal that
gravity is thermodynamic in nature.

In all cases the supergravity analyses were formulated in terms of classical solutions, but any such mapping must also extend to
the quantum level. Such an extension for the LLM system was proposed in \cite{usbig}, and in this note we apply the methods
developed in that paper to the D1-D5 system.

We propose a `metric' operator in the CFT: an operator whose eigenstates are dual to semi-classical geometries via the mapping
given in \cite{jan}. The states that fail to be eigenstates, however, cannot be mapped to spacetimes with unique metrics.

We also analyse how the data characterising the field theory state shows up in the asymptotic form of the spacetime metric. We
find the data to be arranged into a set of multipoles, the first of which was already considered in \cite{jan,jan2} as the dipole
operator. We also find that certain terms in the metric only show up if the CFT dual state is a superposition of basis states,
and demonstrate the measurability of these interference effects. Both of these results are highly analogous to what was found for
the LLM geometries in \cite{us,usbig}.

Finally, we point out that the thermal ensemble, consisting of a sum over all states with the total twist $N$ fixed using a
lagrange multiplier $\beta$, is not an eigenstate of the metric operator due to the large fluctuations inherent in the ensemble.
This is again highly analogous to what was found for the LLM case in \cite{pettorino}, but we show that the method used there to
restrict the ensemble is incapable of sufficiently constraining the ensemble in the D1-D5 case.

The paper is structured as follows. In section \ref{review} we present a brief review of the D1-D5 system and the map proposed in
\cite{jan}. In section \ref{sec:asymp1} we construct the asymptotic expansion of the metric and find a set of multipoles. In
section \ref{metop} we proceed to use these multipoles to motivate our definition of the metric operator, and define
the approximate eigenstates of this operator. In section \ref{sec:asymp2} we consider a more general asymptotic expansion of the
metric and find the terms due to interference between basis states. In section \ref{sec:thermal} we consider the thermal
ensemble, and we conclude in section \ref{discussion} with some comments.

\section{Review}
\label{review} We begin by briefly reviewing the D1-D5 system; for a more comprehensive review the reader is referred to
\cite{luninmathur,mathur,lunin,vijaymasaki,luninorig,orig2,jansugra}. The D1-D5 CFT, which is dual to type IIB string theory on $\textrm{AdS}_3 \times
\textrm{S}^3 \times \textrm{T}^4$, is a marginal deformation of the ($1+1$)-dimensional orbifold sigma model with target space
\begin{equation}
\mathcal{M}_0 =  \left( \textrm{T}^4 \right)^N \! \! / \textrm{S}_N,
\end{equation}
where $N$ is related to the AdS scale and $\textrm{S}_N$ is the permutation group. This duality arises as the decoupling limit of
type IIB string theory on $\textrm{M}^{1,4} \times \textrm{S}^1 \times \textrm{T}^4$ with $N_1$ D1-branes wrapping the
$\textrm{S}^1$ and $N_5$ D5-branes wrapping $\textrm{S}^1 \times \textrm{T}^4$, where the parameters are related by $N=N_1N_5$.

\paragraph{Gravity solutions:} The microstate geometries of the D1-D5 system are well known and can be written as
\begin{eqnarray}
ds^2 & = & \frac{1}{\sqrt{f_1f_5}} [ -(dt + A)^2 + (dy + B)^2 ] + \sqrt{f_1f_5} d\vec{x}^2 + \sqrt{\frac{f_1}{f_5}} d\vec{z}^2, \\
& & e^{2\Phi} = \frac{f_1}{f_5}, \quad C = \frac{1}{f_1} (dt+A) \wedge (dy+B) + \mathcal{C}, \\
& & dB = \ast_4dA, \quad d\mathcal{C} = - \ast_4 df_5, \\
& & f_5 = 1 + \frac{Q_5}{L} \int_0^L \frac{ds}{|\vec{x} - \vec{F}(s) |^2}, \\
& & f_1 = 1 + \frac{Q_5}{L} \int_0^L \frac{|\vec{F}'(s)|^2 ds}{|\vec{x} - \vec{F}(s) |^2}, \\
& & A_i = \frac{Q_5}{L} \int_0^L \frac{F_i'(s) ds}{|\vec{x} - \vec{F}(s) |^2}.
\end{eqnarray}
Here $y$ and $\vec{z}$ parametrize the $\textrm{S}^1$ and $\textrm{T}^4$ respectively. The coordinate radius of the
$\textrm{S}^1$ is $R$, while the coordinate volume of the $\textrm{T}^4$ is $V_4$. The charges $Q_1$ and $Q_5$ are related to
$N_1$ and $N_5$ by
\begin{equation}
Q_5 = g_s  N_5, \quad Q_1 = \frac{g_s }{V_4} N_1. \Label{Qs}
\end{equation}

All these solutions are parametrized in terms of a closed curve $\vec{F}(s)$ in $\mathbb{R}^4$, which we expand as a Fourier
series as
\begin{equation}
\vec{F}(s) = \mu \sum_{\stackrel{k=-\infty}{k\neq 0}}^{\infty} \frac{1}{\sqrt{2|k|}} \vec{d}_k e^{i\frac{2\pi k}{L}s},
\Label{fourier}
\end{equation}
where $s$ ranges from 0 to $L$ and $\vec{d}_k = (d_k^1,d_k^2,d_k^3,d_k^4) = \vec{d}_{-k}^*$. Note that the fermionic oscillations
as well as oscillations on the $\textrm{T}^4$ have been omitted, as we are only interested in fluctuations in the $\mathbb{R}^4$.
Additionally,
\begin{equation}
\mu = \frac{g_s}{R\sqrt{V_4}}. \Label{mu}
\end{equation}
The parameter $L$ satisfies
\begin{equation}
LR = 2\pi Q_5, \Label{L}
\end{equation}
and due to fixed length of the original string there is an additional constraint
\begin{equation}
Q_1 = \frac{Q_5}{L} \int_0^L |\vec{F}'(s)|^2 ds.
\end{equation}

It was shown in \cite{rychkov} that the space of classical solutions can be quantized to yield a finite number of quantum states.
The quantized system is given by\footnote{This is the only time we use $\vec{d}_k$'s as operators. Our notation is such that
$d_k$'s are complex numbers, while $c_k, \, c_k^{\dag}$ denote annihilation and creation operators.}
\begin{eqnarray}
& & [ d_k^a,d_l^b ] = \delta^{ab} \delta_{kl}, \\
& & \langle \int_0^L :| \vec{F}'(s)|^2: ds \rangle = \frac{(2\pi)^2 \mu^2 N}{L}, \\
& & N=N_1N_5 = \sum_{k=1}^{\infty} k \langle \vec{d}_k^{\, \dag} \cdot \vec{d}_k \rangle.
\end{eqnarray}

\paragraph{Field theory states:} The Ramond ground states of the CFT are in one to one correspondence with states at level $N$ of a Fock
space of a system composed of 8 bosonic and 8 fermionic oscillators.  We shall retain only four of these oscillators; the bosonic
ones that correspond to fluctuations in the transverse $\mathbb{R}^4$. Thus a basis for the states can be written as
\begin{equation}
| \{N_k\} \rangle = \prod_{k=1}^{\infty} \prod_{a=1}^4 \frac{1}{\sqrt{N_k^a!}} \left( c_k^{a\dag} \right)^{N_k^a} |0\rangle,
\quad \textrm{with } \sum_{k=1}^{\infty} \sum_{a=1}^4 kN_k^a = N. \Label{basis}
\end{equation}
For convenience we shall write $\vec{c}_k^{\, \, \dag} \equiv \vec{c}_{-k}$ for positive $k$, so that the notation $\vec{c}$
includes both the creation and annihilation operators.  It was proposed in \cite{jan} to associate a phase space density
$f(\vec{d})$ to each state $|\psi\rangle$ by
\begin{equation}
f_{\psi}(\vec{d}) = \frac{ \langle 0 | e^{\sum_{k=1}^{\infty} \vec{d}_k \cdot \vec{c}_k} | \psi \rangle \langle \psi |
e^{\sum_{k=1}^{\infty} \vec{d}_k^* \cdot \vec{c}_k^{\dag}} | 0 \rangle}{ \langle 0 | e^{\sum_{k=1}^{\infty} \vec{d}_k \cdot
\vec{c}_k}  e^{\sum_{k=1}^{\infty} \vec{d}_k^* \cdot \vec{c}_k^{\dag}} | 0 \rangle}. \Label{eq:f}
\end{equation}
It can be shown that this distribution function corresponds to anti-normal ordering prescription in the quantum system, and can
be used to compute expectation values of anti-normal ordered operators as
\begin{equation}
\int_{\vec{d}} f_{\psi}(\vec{d}) \, g(\vec{d}) = \langle \psi | :g(\vec{c}):_A | \psi \rangle.
\end{equation}
Also, the distribution corresponding to the basis state (\ref{basis}) can be easily computed and gives
\begin{equation}
f_{\{N_k^a\} } (\vec{d}) = \prod_{k=1}^{\infty} \prod_{a=1}^4 e^{-d_k^a d_{-k}^a} \frac{\left(d_k^a d_{-k}^a
\right)^{N_k^a}}{N_k^a!}. \Label{fbasis}
\end{equation}

In addition to this basis, we will often find it useful to work with coherent states. These can be defined as
\begin{equation}
|\{\vec{\tilde{d}} \} \rangle = e^{-\frac{ \vec{\tilde{d}}_k \cdot \vec{\tilde{d}}_k^{*}}{2}} \textrm{P}_N
e^{\vec{\tilde{d}}_k^{*} \cdot \vec{c}_k^{\dag}} | 0 \rangle, \Label{coherent}
\end{equation}
where $\vec{\tilde{d}}_k \in \mathbb{C}^4$ for all $k$, and $\textrm{P}_N$ is a projection operator to the twist $N$ subspace of
the Fock space. Note that we are suppressing the sums over $k$ in the exponents, and that this definition differs from the
definition in \cite{jan} by a normalization factor. With this definition one finds the corresponding distribution to be
\begin{equation}
f_{\vec{\tilde{d}}}(\vec{d}) = \prod_{k=1}^{\infty} e^{-|\vec{d}_k - \vec{\tilde{d}}_k|^2} + \mathcal{O}(\frac{1}{N}),
\end{equation}
where the subleading correction arises because of the projection operator $\textrm{P}_N$, and will vanish in the $N \to \infty$
limit.

Using this distribution, it was proposed in \cite{jan} that the microstate geometry dual to state $|\Psi \rangle$ should be given
by
\begin{eqnarray}
f_5 & = & 1 + \frac{Q_5}{L}\mathcal{N} \int_{\vec{d}}  \int_0^L \frac{f_{\Psi}(\vec{d})\, ds}{|\vec{x} - \vec{F}(s) |^2}, \Label{quantumf5} \\
f_1 & = & 1 + \frac{Q_5}{L}\mathcal{N} \int_{\vec{d}}  \int_0^L \frac{f_{\Psi}(\vec{d})\, |\vec{F}'(s)|^2 ds}{|\vec{x} - \vec{F}(s) |^2}, \Label{quantumf1} \\
A_i & = & \frac{Q_5}{L} \mathcal{N} \int_{\vec{d}} \int_0^L \frac{f_{\Psi}(\vec{d}) \, F_i'(s) ds}{|\vec{x} - \vec{F}(s) |^2},
\Label{quantuma}
\end{eqnarray}
where the normalization factor is
\begin{equation}
\mathcal{N}^{-1} = \int_{\vec{d}} f_{\Psi}(\vec{d}).
\end{equation}

This is a mapping from a quantum system to a set of semiclassical geometries, and we shall see in section \ref{metop} that it
shouldn't be applied to an arbitrary state, or more generally to an arbitrary density matrix, as this may yield unphysical
spacetimes. In section \ref{metop} we propose a metric operator in the CFT, the eigenstates of which can be associated to
microstate geometries using the prescription above.

\section{Asymptotic expansion of a basis state}
\label{sec:asymp1} We wish to determine how the microstate geometry (\ref{quantumf5},\ref{quantumf1},\ref{quantuma})
corresponding to a given basis state (\ref{basis}) appears to an asymptotic observer. To accomplish this, we shall expand $f_5$,
given by (\ref{quantumf5}), as a power series in the inverse radial coordinate $\frac{1}{r}$. For completeness, we also compute
the expansion of $f_1$ in appendix \ref{sec:f1exp}. For $r \gg |\vec{F}(s)|$ we can expand
\begin{equation}
|\vec{x} - \vec{F}(s)|^{-2} = r^{-2} \left(1 - \frac{2\vec{r} \cdot \vec{F}(s) - |\vec{F}(s)|^2}{r^2} \right)^{-1} =
\frac{1}{r^2} \sum_{n=0}^{\infty} \left( \frac{2\vec{r} \cdot \vec{F}(s) - |\vec{F}(s)|^2}{r^2} \right)^n.
\end{equation}
Plugging this into (\ref{quantumf5}) and expanding the binomial we get
\begin{equation}
f_5 = 1 + \frac{Q_5}{L} \mathcal{N} \frac{1}{r^2} \sum_{n=0}^{\infty} \frac{1}{r^{2n}} \sum_{p=0}^n \binom{n}{p} (-1)^p 2^{n-p}
r^{n-p} \int_0^L \int_{\vec{d}} f(\vec{d}) (\vec{e}\cdot \vec{F}(s))^{n-p} |\vec{F}(s)|^{2p},
\end{equation}
where $\vec{r} \equiv r \vec{e}$ and $|\vec{e}|^2 = 1$. To make the powers of $\frac{1}{r}$ more explicit, we define a new
summation index $l \equiv n+p$, which runs from 0 to infinity. Eliminating $n$, we see that $p$ now runs from 0 to $\left[
\frac{l}{2} \right]$. To make the integral more explicit, we also eliminate $\vec{F}(s)$ using (\ref{fourier}) . This gives
\begin{eqnarray}
f_5 & = & 1 + \frac{Q_5}{r^2} \sum_{l=0}^{\infty} \left(\frac{\mu}{r}\right)^l \sum_{p=0}^{\left[ \frac{l}{2} \right]} (-1)^p
2^{\frac{l}{2}-2p} \binom{l-p}{p} \! \! \! \! \sum_{\stackrel{\stackrel{k_1,\ldots,k_p}{l_1,\ldots,l_p}}{m_1,\ldots,m_{l-2p}}} \!
\! \! \!  \frac{\delta(\sum_i (k_i+l_i) + \sum_j m_j)}{\sqrt{|\prod_i k_il_i \prod_j m_j|}} \cdot \nonumber \\ & \cdot &
\mathcal{N} \int_{\vec{d}} \prod_{s=1}^{\infty} \prod_{a=1}^4 e^{-d_s^ad_{s}^{a*}} \left( d_s^a d_s^{a*} \right)^{N_s^a}
\prod_{i=1}^p \left( \vec{d}_{k_i}\cdot \vec{d}_{l_i} \right)\prod_{j=1}^{l-2p} \left( \vec{e} \cdot \vec{d}_{m_j}\right),
\end{eqnarray}
where the integral over $s$ gave rise to the Kronecker delta. The integral can only be non-zero when the number of $\vec{d}$'s is
even, so we can write $l \equiv 2n$, which gives
\begin{eqnarray}
f_5 & = & 1 + \frac{Q_5}{r^2} \sum_{n=0}^{\infty} \left(\frac{\mu}{r}\right)^{2n} \sum_{p=0}^n (-1)^p 2^{n-2p} \binom{2n-p}{p} \!
\! \! \! \sum_{\stackrel{\stackrel{k_1,\ldots,k_p}{l_1,\ldots,l_p}}{m_1,\ldots,m_{2(n-p)}}} \! \! \! \! \frac{\delta(\sum_i
(k_i+l_i) + \sum_j m_j)}{\sqrt{|\prod_i k_il_i \prod_j m_j|}} \cdot \nonumber \\ & \cdot & \mathcal{N} \int_{\vec{d}}
\prod_{s=1}^{\infty} \prod_{a=1}^4 e^{-d_s^ad_{s}^{a*}} \left( d_s^a d_{s}^{a*} \right)^{N_s^a} \prod_{i=1}^p \left(
\vec{d}_{k_i}\cdot \vec{d}_{l_i} \right)\prod_{j=1}^{2(n-p)} \left( \vec{e} \cdot \vec{d}_{m_j}\right). \Label{basisexp}
\end{eqnarray}
In the above all the remaining integrals are gaussian. However, the combinatorics of the indices $k_i, \, l_i$ and $m_j$ quickly
become untractable and we have been unable to find a closed form expression for the $n^{\textrm{th}}$ level of the expansion. In
appendix \ref{combi} we present a procedure that can in principle be used compute any given order, though it quickly becomes very
tedious for higher orders.

Lacking a general closed form for the expansion, we can at least compute the first few nontrivial orders. For simplicity, we also
take the occupation numbers to be independent of direction in the $\mathbb{R}^4$, i.e. $N_k^a = N_k$. As shown in the appendix,
we get
\begin{equation}
f_5 = 1 + \frac{Q_5}{r^2} - 12 \frac{Q_5 \mu^4}{r^6}  M_2 + 40 \frac{Q_5 \mu^6}{r^8} M_3 + \mathcal{O}(\frac{1}{r^{10}}),
\Label{asympfirst}
\end{equation}
where we have defined the multipoles
\begin{equation}
M_k = \sum_{m=1}^{\infty} \frac{\left(N_m\right)^k}{m^k}. \Label{multipoles}
\end{equation}
As argued in the appendix, the multipole $M_k$ will first appear in the coefficient of $\frac{1}{r^{2k+2}}$ in the expansion. The
measurability of these higher order terms depends on how they scale as $N$ is taken to infinity. The average occupation numbers
are given by Bose--Einstein statistics, a fact we shall show in section \ref{sec:thermal}; for now we just take this as a given
and find
\begin{equation}
\langle M_k \rangle = \sum_{m=1}^{\infty} \frac{1}{m^k} \frac{1}{(e^{\beta m} - 1)^k} \approx \sum_{m=1}^{\infty}
\frac{1}{m^{2k}} \frac{1}{\beta^k} \sim N^{\frac{k}{2}},
\end{equation}
where the inverse temperature scales as $\beta \propto N^{-\frac{1}{2}}$.  We also know that $r \propto N^{\frac{1}{4}}$ and $Q_5
\propto \sqrt{N}$, from which it follows that the combination $\frac{Q_5 M_k}{r^{2k+2}}$ is remains finite in the limit $N \to
\infty$, and therefore the higher order terms in the expansion are measurable for an observer that can make measurements with
sufficient precision. Since $f_5$ appears directly in the metric, an asymptotic observer can measure these multipoles and
retrieve some data about the CFT state.

To be more precise, an asymptotic observer can measure the multipole $M_k$ by measuring the $(2k+2)^{th}$ derivative of the
metric, or a suitable invariant composed of the derivatives. If such a measurement is made with a machine of finite spatial size
$\lambda$, the resolution of the machine must be at least $\lambda / 2k$. Since any measurement is bounded by the Planck scale,
this gives a condition
\begin{equation}
\frac{\lambda}{2k} > l_p^{(6)},
\end{equation}
where the six-dimensional Planck length is defined in terms of the 6D Newton's constant and the 6D string coupling as
$(l_p^{(6)})^4 = G_6 = g_6^2$. If the size of the measurement apparatus is $\lambda = \gamma R_{\textrm{AdS}_3}$, we get
\begin{equation}
k \lesssim \frac{\gamma R_{\textrm{AdS}_3}}{\sqrt{g_6}} = \gamma \frac{\sqrt{g_6} N^{\frac{1}{4}}}{\sqrt{g_6}} = \gamma
N^{\frac{1}{4}}.
\end{equation}
This gives a limit to how much CFT data an asymptotic observer with sufficient ingenuity can measure. However, this bound is very
likely to be too generous; measuring multipoles of order $k \sim N^{\frac{1}{4}}$ involves high energies, the backreaction of
which on the geometry cannot be ignored. Thus it is no longer sufficient to work in the $\frac{1}{2}$-BPS sector without taking
into account the $\alpha'$ and $g_s$ corrections, which are likely to impose a tighter bound on how many multipoles are
measurable. In this note we will not attempt to analyse this in more detail.

\section{The metric operator}
\label{metop}  We shall now explain our earlier statement that the map (\ref{quantumf5},\ref{quantumf1},\ref{quantuma}) does not
extent to all the states $|\Psi\rangle$ in the Hilbert space.  Consider a superposition of two very different states, say
\begin{equation}
|\Psi \rangle = \frac{1}{\sqrt{2}} |\psi_1 \rangle + \frac{1}{\sqrt{2}} |\psi_2 \rangle, \quad \textrm{with} \quad | \psi_1
\rangle = \prod_{a=1}^4 \frac{1}{\sqrt{(N/4)!}}\left( c_1^{a\dag} \right)^{\frac{N}{4}} |0\rangle, \quad \textrm{and} \quad
|\psi_2 \rangle = \prod_{a=1}^4 c_{N/4}^{a\dag} |0\rangle. \Label{badsuperpos}
\end{equation}
Note that neither of these states is typical in any sense, but they serve to illustrate the issue; we will deal with the full
thermal ensemble of states in section \ref{sec:thermal}. We immediately find the multipoles $M_k$ in these states as\footnote{Due
to the non-typicality, these don't scale as $N^{\frac{k}{2}}$ like they would in a typical state. Indeed, $\psi_1$ has the
maximal possible multipoles, while $\psi_2$ has the smallest possible multipoles.}
\begin{equation}
M_k^{\psi_1} = \left( \frac{N}{4} \right)^k, \quad M_k^{\psi_2} = \frac{1}{\left( \frac{N}{4} \right)^k}.
\end{equation}
Since (\ref{eq:f}) and (\ref{quantumf5}) are linear\footnote{The density matrix for $\Psi$ will have cross terms $|\psi_1 \rangle
\langle \psi_2 |$ and  $|\psi_2 \rangle \langle \psi_1 |$. However, we shall show in section \ref{sec:asymp2} that these will
have minimal contribution to the phase space distribution and will not affect the multipoles. Therefore the distribution is the
sum of the distributions of $\psi_1$ and $\psi_2$.} in the density matrix, the multipoles of the state $|\Psi \rangle$ are given
by $M_k^{\Psi} = \frac{1}{2} (M_k^{\psi_1} + M_k^{\psi_2})$, which is very different from both $M_k^{\psi_1}$ and $M_k^{\psi_2}$.
This is not problematic from the CFT point of view, but the spacetime interpretation presents a problem. As soon as an observer
measures any of the multipoles in the spacetime, standard measurement theory arguments state that the universe is projected into either of the two states $\psi_1$ or $\psi_2$.
But the three geometries differ from each other at scales which are easily measurable and therefore `jumping' between these
metrics based on one measurement is not physically acceptable. Because of this problem we need to develop a criterion that
establishes when a state can be mapped into a microstate geometry using (\ref{quantumf5},\ref{quantumf1},\ref{quantuma}), and
when it's not reasonable to associate a semiclassical metric to a state in the CFT.

\subsection{The metric operator and eigenstates}
\label{2nd} We shall now define the general multipole operator\footnote{The idea of using a formalism like this to determine which
states can be mapped to unique semiclassical geometries was first used in the setting of $\frac{1}{2}$-BPS sector of
$\mathcal{N}=4$ SU($N$) Yang-Mills in \cite{usbig}.} as
\begin{equation}
\hat{M}(k) \equiv \sum_{m=1}^{\infty} \frac{1}{m^k}\hat{N}_m^k = \sum_{m=1}^{\infty} \frac{1}{m^k} \left( c_m^{\dag} c_m
\right)^k,
\end{equation}
which is simply the quantum version of (\ref{multipoles}). Note that we are suppressing the $\mathbb{R}^4$ indices.

Next we need to define what we mean by approximate eigenstates of the operator $\hat{M}(k)$. From the definition it is clear that
the only exact eigenstates are the basis states (\ref{basis}), while any superposition is necessarily not an eigenstate. This is
too restricting; rather we wish to introduce a coarse graining to correspond to the limited measurement precision of an observer.
To do this, for an arbitrary state $|\Psi\rangle $ we  introduce the functional
\begin{equation}
E[M(k)] = \textrm{Tr} \left[ \hat{\rho}_{\Psi} \left( \hat{M}(k) - M(k) \right)^2 \right], \Label{functional}
\end{equation}
and we shall call the function that minimizes this functional $M_{\Psi}(k)$. Thus armed, we say that $|\Psi \rangle$ is an
eigenstate of $\hat{M}(k)$ with eigenvalue function $M_{\Psi}(k)$ and accuracies $\{ \epsilon_k \}$, iff
\begin{equation}
\frac{\sqrt{E[M_{\Psi}(k)]}}{M_{\Psi}(k)} < \epsilon_k, \quad \textrm{for all } k. \Label{criterion}
\end{equation}
Note that if $|\Psi \rangle = |\{ N_k\} \rangle$ is a basis state, then $E[M_{\Psi}(k)] = 0$, with $M_{\Psi}(k)$ given by
(\ref{multipoles}), and (\ref{criterion}) is trivially satisfied.

With this definition, we are finally in a position to state our proposal in a definite form: \vspace{0.2cm}

{\it The states in the CFT that have good dual description in terms of a unique metric are the ones that are approximate eigenstates
of the operator $\hat{M}(k)$.} \vspace{0.2cm}

In this sense we can also call $\hat{M}(k)$ a `metric' operator: its eigenstates are the only ones that can be mapped to
semi-classical spacetimes with unique metrics, and its eigenvalue functions specify the multipoles present in the asymptotic
expansion of the metric and allow an observer to reconstruct the metric up to some measurement precision.

\subsection{Explicit example}
Before closing this section, we wish to illustrate this formalism by considering an explicit example. We choose the state to be a
superposition of two basis states: \mbox{$|\Psi \rangle = \frac{1}{\sqrt{2}} (|\{ N_{m1} \} \rangle + |\{ N_{m2} \} \rangle)$.}
The expectation values in (\ref{functional}) are easily evaluated and yield
\begin{eqnarray}
\langle \Psi | \hat{M}(k) | \Psi \rangle & = & \frac{1}{2} \left( \sum_{m=1}^{\infty} \frac{N_{m1}^k}{m^k} +
\sum_{m=1}^{\infty} \frac{N_{m2}^k}{m^k} \right),  \\
\langle \Psi | \hat{M}(k)^2 | \Psi \rangle & = & \sum_{m,n=1}^{\infty} \frac{1}{m^kn^k} \langle\Psi | \hat{N}_m^k \hat{N}_n^k |
\Psi \rangle =  \frac{1}{2} \left( \left(\sum_{m=1}^{\infty} \frac{N_{m1}^k}{m^k}\right)^2  + \left(\sum_{m=1}^{\infty}
\frac{N_{m2}^k}{m^k} \right) ^2\right). \nonumber
\end{eqnarray}
Plugging these into the functional (\ref{criterion}), we can write it as
\begin{equation}
E[M(k)] = \left( M(k) - \frac{1}{2} \sum_{m=1}^{\infty} \frac{N_{m1}^k + N_{m2}^k}{m^k} \right)^2 + \frac{1}{4} \left(
\sum_{m=1}^{\infty} \frac{N_{m1}^k}{m^k} - \sum_{m=1}^{\infty} \frac{N_{m2}^k}{m^k} \right)^2.
\end{equation}
This is minimized by choosing $M_{\Psi}(k) = \frac{1}{2} \sum \frac{N_{m1}^k + N_{m2}^k}{m^k} = \frac{1}{2} (M_{k,\{ N_{m1}\} } +
M_{k,\{ N_{m2}\} }) $, i.e. average of the multipoles of the two states. However, the functional never vanishes and the condition
(\ref{criterion}) can be written as
\begin{equation}
\frac{ |M_{k,\{ N_{m1}\} } - M_{k,\{ N_{m1}\} } | }{M_{k,\{ N_{m1}\} } + M_{k,\{ N_{m2}\} }} < \epsilon_k,
\end{equation}
which gives a condition for how much the multipoles of the two states can differ if $|\Psi\rangle$ is to be an eigenstate with
accuracy $\epsilon_k$. For the superposition considered at the beginning of this section, (\ref{badsuperpos}), the ratio above is
of order one, and therefore this state is far from being an eigenstate.

\section{More asymptotic expansions}
\label{sec:asymp2} We now wish to find the asymptotic expansion for a general state in the theory, rather than just for basis
states. Of course, for any state we need to check that it is an approximate eigenstate of $\hat{M}(k)$ before we can trust this
expansion. A general superposition is given by
\begin{equation}
|\psi \rangle = \sum_w \alpha_w \prod_{k=1}^{\infty} \prod_{a=1}^4 \frac{\left( c_k^{a \dag}
\right)^{N_k^{a,w}}}{\sqrt{N_k^{a,w}!}} |0\rangle, \quad \textrm{with } \sum_{k=1}^{\infty} \sum_{a=1}^4 k N_k^{a,w} = N \, \,
\forall w, \quad \textrm{and } \sum_w |\alpha_w|^2 =1,
\end{equation}
where $w$ indexes the states in the superposition. The phase space distribution can again be computed, and yields
\begin{eqnarray}
f(\vec{d}) & = & \sum_{w,w'} \alpha_w \alpha_{w'}^* \prod_{k=1}^{\infty} \prod_{a=1}^4 e^{-d_k^ad_{k}^{a*}} \left( d_k^a
\right)^{N_k^{a,w}} \left( d_{k}^{a*} \right)^{N_{k}^{a,w'}} \Label{superf} \\
& = & \sum_{w,w'} \alpha_w \alpha_{w'}^* \prod_{k=1}^{\infty} \prod_{a=1}^4 e^{-(\rho_k^a)^2} (\rho_k^a)^{(N_k^{a,w}+N_k^{a,w'})}
e^{i\phi_k^a (N_k^{a,w}-N_k^{a,w'})}, \nonumber 
\end{eqnarray}
where in the second equality we have switched to polar coordinates. Thus we can see that all angular dependence in the phase
space distribution is due to interference terms between different basis states. Following the recipe laid out in section
\ref{sec:asymp1}, we can expand $f_5$ in $\frac{1}{r}$ to get
\begin{eqnarray}
f_5 & = & 1 + \frac{Q_5}{r^2} \sum_{w,w'} \alpha_w \alpha_{w'}^* \sum_{l=0}^{\infty} \left(\frac{\mu}{r}\right)^l
\sum_{p=0}^{\left[ \frac{l}{2} \right]} (-1)^p 2^{\frac{l}{2}-2p} \binom{l-p}{p} \! \! \! \!
\sum_{\stackrel{\stackrel{k_1,\ldots,k_p}{l_1,\ldots,l_p}}{m_1,\ldots,m_{l-2p}}} \! \! \! \! \frac{\delta(\sum_i (k_i+l_i) +
\sum_j m_j)}{\sqrt{|\prod_i k_il_i \prod_j m_j|}} \cdot \nonumber \\ 
& \cdot & \mathcal{N}  \int_{\vec{d}} \prod_{k=1}^{\infty}
\prod_{a=1}^4 e^{-d_k^ad_{k}^{a*}} \left( d_k^a \right)^{N_k^{a,w}} \left( d_{k}^{a*} \right)^{N_{k}^{a,w'}} \prod_{i=1}^p \left(
\vec{d}_{k_i}\cdot \vec{d}_{l_i} \right)\prod_{j=1}^{l-2p} \left( \vec{e} \cdot \vec{d}_{m_j}\right). \Label{superexp}
\end{eqnarray}
Though analyzing this in detail is untractable, we can still make some interesting observations. Since all the terms in the phase
space distribution (\ref{superf}) do not in general have an even number of $d$'s, we see that the summation index $l$ does not
need to be even anymore, and thus the expansion now has terms that are odd in $\frac{1}{r}$. The origin of these terms is
completely due to interference between basis states.

\subsection{Expansion for coherent states}
Analyzing the measurability of the odd terms in (\ref{superexp}) is difficult when working in the basis of eigenstates of
excitation numbers. However, using coherent states we can explicitly show that these terms can be measurable. The phase space
distribution corresponding to a coherent state was written down in (\ref{coherent}), and using it we can once again expand
(\ref{quantumf5}) to get
\begin{eqnarray}
f_5 & = & 1 + \frac{Q_5}{r^2} + 4 \frac{Q_5 \mu^2}{r^4} \sum_{m=1}^{\infty} \frac{1}{m} \left[ (\vec{\tilde{d}}_m \cdot \vec{e})
(\vec{\tilde{d}}_{-m} \cdot \vec{e}) - (\vec{\tilde{d}}_m \cdot \vec{\tilde{d}}_{-m}) \right] + \Label{coherentexp} \\
& & + \sqrt{2} \frac{Q_5\mu^3}{r^5}  \sum_{k,l,m} \frac{\delta(k+l+m)}{\sqrt{|klm|}} \left[ 2(\vec{\tilde{d}}_m \cdot
\vec{e})(\vec{\tilde{d}}_k \cdot \vec{e})(\vec{\tilde{d}}_l \cdot \vec{e}) -  (\vec{\tilde{d}}_k \cdot \vec{\tilde{d}}_l)(\vec{e}
\cdot \vec{\tilde{d}}_m) \right] + \mathcal{O}(\frac{1}{r^6}). \nonumber
\end{eqnarray}
To complete the analysis, we need to show that these odd terms are measurable and that coherent states are approximate
eigenstates of $\hat{M}(k)$ and therefore it is sensible to associate semiclassical geometries to them.

\paragraph{Measurability:}
We need to determine how the $\vec{\tilde{d}}$'s scale as a function of $N$. To do this, we compute the overlap between the
coherent state and an arbitrary basis state. This can be done using (\ref{basis}) and (\ref{coherent}), and gives
\begin{equation}
\langle \{ N_k\} | \vec{\tilde{d}} \rangle = \prod_{k=1}^{\infty} e^{-\frac{|d_k|^2}{2}} \frac{(d_k^*)^{N_k}}{\sqrt{N_k!}}.
\Label{overlap}
\end{equation}
To determine which basis state has the largest overlap with the coherent state, we maximize the norm of (\ref{overlap}) and find
\begin{equation}
|\vec{\tilde{d}}_k| = \sqrt{N_k}. \Label{dscaling}
\end{equation}
We want the $N \to \infty$ limit to be one that leaves inner products like (\ref{overlap}) unchanged; hence (\ref{dscaling})
tells us the scaling of $\vec{\tilde{d}}_k$. For states near the typical state, $N_k \propto \sqrt{N}$ for small $k$, and
therefore we see that the terms in the expansion remain fixed as $N$ is scaled\footnote{Remember that $r$ scales as $N^{1/4}$ and
$Q_5$ as $\sqrt{N}$}. This is enough to show that the effects of interference remain observable, even in the $N \to \infty$
limit.

\paragraph{Eigenstates:}
Finally, we need to show that the coherent states are approximate eigenstates of $\hat{M}(k)$. The expectation values of
$\hat{M}(k)$ and $\hat{M}(k)^2$ are
\begin{eqnarray}
\textrm{Tr}(\hat{\rho} \hat{M}(k)) & = & \sum_{m=1}^{\infty} \frac{1}{m^k} \sum_{\{ N_p\} } |\langle \{ N_p\} | \vec{\tilde{d}}
\rangle |^2 \langle \{ N_p\} | N_m^k | \{ N_p \} \rangle \nonumber \\
& = & \sum_{m=1}^{\infty} \frac{1}{m^k} e^{-|d_m|^2} \sum_{N_m=0}^{\infty} \frac{|d_m|^{2N_m}}{N_m!} (N_m)^k, \\
\textrm{Tr}(\hat{\rho} \hat{M}(k)^2) & = & \sum_{m,n=1}^{\infty} \frac{1}{m^kn^k} \sum_{\{ N_p\} } |\langle \{ N_p\} |
\vec{\tilde{d}} \rangle |^2 \langle \{ N_p\} | N_m^kN_n^k | \{ N_p \} \rangle \nonumber \\
& = & \left( \sum_{m=1}^{\infty} \frac{e^{-|d_m|^2}}{m^k} \sum_{N_m=0}^{\infty} \frac{|d_m|^{2N_m}}{N_m!}N_m^k \right)^2 +
\nonumber \\ & + & \sum_{m=1}^{\infty} \frac{e^{-|d_m|^2}}{m^{2k}} \sum_{N_m=0}^{\infty} \frac{|d_m|^{2N_m}}{N_m!} N_m^{2k} -
\sum_{m=1}^{\infty} \frac{e^{-2|d_m|^2}}{m^{2k}} \left( \sum_{N_m=0}^{\infty} \frac{|d_m|^{2N_m}}{N_m!} N_m^k \right)^2.
\end{eqnarray}
Using these, one shows that the functional $E[M(k)]$ in (\ref{functional}) can be written as
\begin{eqnarray}
\textrm{Tr}(\hat{\rho}(\hat{M}(k) - M(k))^2) & = & \left( M(k) - \sum_{m=1}^{\infty} \frac{1}{m^k} e^{-|d_m|^2}
\sum_{n=0}^{\infty} \frac{|d_m|^{2n}}{n!} n^k \right)^2 + \nonumber \\
& + & \sum_{m=1}^{\infty} \frac{e^{-|d_m|^2}}{m^{2k}} \left[ \sum_{n=0}^{\infty} \frac{|d_m|^{2n}}{n!}n^{2k} - e^{-|d_m|^2}
\left( \sum_{n=0}^{\infty} \frac{|d_m|^{2n}}{n!}n^k \right)^2 \right].
\end{eqnarray}
The function $M_{\tilde{d}}(k)$ is again chosen such that the first square vanishes. Thus it remains to show that the remaining
terms yield a negligible contribution. To do this, we note that the sums appearing in the the expression above can be computed as
\begin{equation}
e^{-r} \sum_{n=0}^{\infty} \frac{r^{n}}{n!} n^k = e^{-r} (r\partial_r)^k \sum_{n=0}^{\infty} \frac{r^n}{n!} = e^{-r}
(r\partial_r)^k e^{r} = \textrm{Polynomial of order $k$ in $r$}.
\end{equation}
Using this and writing $|d_m|^2 = r_m$, we can write the ratio (\ref{criterion}) as
\begin{equation}
\frac{\sqrt{E[M_{\tilde{d}}(k)]}}{M_{\tilde{d}}(k)} = \frac{ \sqrt{\sum_{m=1}^{\infty} \frac{e^{-r_m}}{m^{2k}} \left[
\sum_{n=0}^{\infty} \frac{r_m^{n}}{n!}n^{2k} - e^{-r_m} \left( \sum_{n=0}^{\infty} \frac{r_m^{n}}{n!}n^k \right)^2 \right] }}{
\sum_{m=1}^{\infty} \frac{1}{m^k} e^{-r_m} \sum_{n=0}^{\infty} \frac{r_m^{n}}{n!} n^k }.
\end{equation}
The denominator is a polynomial of order $k$ in $r_m$, while in the numerator the highest order in $r_m$ cancels and one is left
with a square root of a polynomial of order $2k-1$ in $r_m$. Using the scaling (\ref{dscaling}) we then see
\begin{equation}
\frac{\sqrt{E[M_{\tilde{d}}(k)]}}{M_{\tilde{d}}(k)} \sim \frac{ \sqrt{|d_m|^{4k-2}}}{|d_m|^{2k}} \sim \frac{1}{|d_m|} \sim
\frac{1}{N^{\frac{1}{4}}},
\end{equation}
showing that for large $N$ this is suppressed and the coherent state is an approximate eigenstate to a high precision.

\section{The canonical ensemble}
\label{sec:thermal} Explicit computations in the microcanonical ensemble involving only states of a fixed total twist $N$ can be
complicated. One often used method of circumventing this is to work in a canonical ensemble, fixing the total twist to equal $N$
using a Lagrange multiplier. However, we shall show that this ensemble is not well suited for use with the mapping
(\ref{quantumf5},\ref{quantumf1},\ref{quantuma}), and this is possibly the reason why in \cite{jan} a non-standard entropy was
found for the $M=0$ BTZ black hole.

 In the canonical ensemble the thermal density matrix can be written as
\begin{equation}
\hat{\rho} = \sum_{ \{ N_k \} } \frac{e^{-\beta \hat{N}} |\{ N_k\} \rangle \langle \{ N_k\} |}{\textrm{Tr}(e^{-\beta \hat{N}})} =
\prod_{k=1}^{\infty} (1-e^{-\beta k}) \sum_{N_k=0}^{\infty} e^{-\beta k N_k} |k,N_k\rangle \langle k,N_k |, \Label{rho}
\end{equation}
where $|k,N_k\rangle = \frac{1}{\sqrt{N_k!}} (c_k^{\dag})^{N_k} |0\rangle$, and $\beta$ has to be fixed by the condition $\langle
\hat{N} \rangle = N$. Note that we're treating all directions as isotropic, and thus suppressing the $\mathbb{R}^4$ index $a$.
The expected occupation numbers and total twist were computed in \cite{jan} to give
\begin{eqnarray}
& & \langle \hat{N}_m \rangle = \textrm{Tr}(\hat{\rho}\hat{N}_m) = (1-e^{-\beta m}) \sum_{N_m=0}^{\infty} N_m e^{-\beta m N_m} = \frac{1}{e^{\beta m} - 1}, \\
& & \langle \hat{N} \rangle = \sum_{m=1}^{\infty} m\langle \hat{N}_m \rangle = \frac{2\pi^2}{3\beta^2}.
\end{eqnarray}
The second equation fixes the inverse temperature
\begin{equation}
\beta = \pi \sqrt{\frac{2}{3N}}. \Label{beta}
\end{equation}
In addition to these we will need the expectation values of higher powers of the occupation numbers. For $\beta m \ll 1$, we can
find them by approximating the sum by an integral, which yields
\begin{equation}
\langle \hat{N}_m^k \rangle = (1-e^{-\beta k}) \sum_{N_m=0}^{\infty} N_m^k e^{-\beta m N_m} \approx (1-e^{-\beta k})
\int_0^{\infty} dN_m N_m^k e^{-\beta m N_m} \approx \frac{k!}{\beta^km^k}. \Label{Nkexp}
\end{equation}

\subsection{Limitations of the canonical ensemble}

There is a problem with using the canonical ensemble with the CFT-to-gravity mapping
(\ref{quantumf5},\ref{quantumf1},\ref{quantuma}), as can be seen by computing the standard deviation to mean ratio of the
occupation numbers\footnote{This looks different from what (\ref{Nkexp}) would give, as this is an exact result. To leading order
(\ref{Nkexp}) will give the same result.}:
\begin{equation}
\frac{\sigma(\hat{N}_k)}{\langle \hat{N}_k \rangle} = \frac{\sqrt{ \langle \hat{N}_k^2 \rangle - \langle \hat{N}_k
\rangle^2}}{\langle \hat{N}_k \rangle} = e^{\frac{\beta k}{2}}.
\end{equation}
This doesn't vanish in the $N\to \infty$ limit, and is an indication that the fluctuations in the occupation numbers are always
large. This doesn't invalidate the ensemble as such, since one can show that the fluctuations in the total twist $\langle \hat{N}
\rangle$ are of the order $N^{-\frac{1}{4}}$ and therefore the ensemble samples only states of twist $N$ to a good degree.
However, in using the CFT-to-gravity mapping the fluctuations in $\hat{N}_k$'s are of paramount importance, as they lead to large
fluctuations in the multipoles $\hat{M}_k$, which in turn lead to superpositions of states of very different metrics, as in the
example at the beginning of section \ref{metop}. Thus, this thermal state should not be mapped to a geometry at all. Indeed, we
can check that this density matrix does not satisfy (\ref{criterion}) and therefore does not pass our criterion.  We can use
(\ref{Nkexp}) to compute the expectation value of $\hat{M}(k)$;
\begin{equation}
\textrm{Tr}(\hat{\rho} \hat{M}(k) ) = \sum_{m=1}^{\infty} \frac{1}{m^k} \textrm{Tr} (\hat{\rho} \hat{N_m^k} ) =
\sum_{m=1}^{\infty} \frac{\langle N_m^k \rangle}{m^k} \approx \sum_{m=1}^{\infty} \frac{k!}{\beta^km^{2k}} = \frac{k!
\zeta(2k)}{\beta^k},
\end{equation}
and the expectation value of the square
\begin{eqnarray}
\textrm{Tr}(\hat{\rho} \hat{M}(k)^2 ) & = & \sum_{m,n=1}^{\infty} \frac{1}{m^kn^k}\langle \hat{N}_m^k\hat{N}_n^k \rangle \approx
\sum_{\stackrel{m,n=1}{m\neq n}}^{\infty} \frac{k!^2}{\beta^{2k}m^{2k}n^{2k}} + \sum_{m=1}^{\infty}
\frac{(2k)!}{\beta^{2k}m^{4k}}  \\
& = & \left( \sum_{m=1}^{\infty} \frac{k!}{\beta^k m^{2k}} \right)^2 + \sum_{m=1}^{\infty} \frac{(2k)!-k!^2}{\beta^{2k} m^{4k}} =
\left( \frac{k! \zeta(2k)}{\beta^k} \right)^2 + \frac{(2k)!-k!^2}{\beta^{2k}}\zeta(4k). \nonumber
\end{eqnarray}
Putting these two results together we can again compute functional (\ref{functional}):
\begin{equation}
E[M(k)] = \left( M(k) - \frac{k! \zeta(2k)}{\beta^k} \right)^2 + \frac{(2k)!-k!^2}{\beta^{2k}}\zeta(4k),
\end{equation}
which is minimized by choosing $M_{\hat{\rho}}(k) = \frac{k! \zeta(2k)}{\beta^k}$. However, the second term will not vanish, and
moreover is not small by any criterion as can be seen by computing the ratio in (\ref{criterion}):
\begin{equation}
\frac{\sqrt{E[M_{\hat{\rho}}(k)]}}{M_{\hat{\rho}}(k)} \approx \sqrt{\frac{(2k)!}{k!^2}-1}  > 1,
\end{equation}
which is greater than any reasonable measurement precision $\epsilon_k$.  Thus the mixed thermal state is not an approximate
eigenstate of $\hat{M}(k)$ and should not be associated to any semi-classical geometry.

\subsection{A restricted canonical ensemble?}
Due to the limitations stated above, we would like to in some way restrict the canonical ensemble in order to curb down the
fluctuations in the multipoles. The most obvious way of doing this would be to fix the first $p$ excitation numbers $N_1,\ldots
,N_p$ to be given by the Bose--Einstein excitation numbers (\ref{Ncusual}), either by hand or using Lagrange multipliers. This
would be in close analogy with what was found in the LLM case in \cite{pettorino}, where one had to restrict the ensemble by
fixing the highest excitation in the system to curb the fluctuations in the ensemble. We shall explore this and other
similarities with the LLM case in the discussion section. Unfortunately, in our case this method fails to sufficiently stabilize
the ensemble, though we feel it is still interesting to present the analysis and investigate why this is so.

Thus we begin by fixing
\begin{equation}
N_{m} \equiv N_c^{(m)} = \frac{1}{e^{\beta m}-1}, \Label{Ncusual}
\end{equation}
so that the the density matrix reduces to
\begin{equation}
\hat{\rho} = |1,N_c^{(1)} \rangle \langle 1,N_c^{(1)} | \otimes \ldots \otimes |p,N_c^{(p)} \rangle \langle p,N_c^{(p)} | \otimes
\left( \prod_{k=p+1}^{\infty} (1-e^{-\beta k}) \sum_{N_k=0}^{\infty} e^{-\beta k N_k} |k,N_k\rangle \langle k,N_k | \right).
\end{equation}
 Using
this density matrix it is clear that the first $p$ excitation numbers do not fluctuate at all, and the fluctuations of the higher
$N_m$'s are as in the unrestricted ensemble. Using the results from the previous subsection it is easy to compute the functional
(\ref{functional}), which gives
\begin{equation}
E[M(k)] = \left[ M(k) - \left( \sum_{m=1}^{p} \frac{(N_c^{(m)})^k}{m^k} + \sum_{m=p+1}^{\infty} \frac{k!}{\beta^k m^{2k}} \right)
\right]^2 + \sum_{m=p+1}^{\infty} \frac{(2k)! - k!^2}{\beta^{2k}m^{4k}}.
\end{equation}
Choosing $M(k)$ to minimize the first square, we can compute the ratio
\begin{equation}
\frac{\sqrt{E[M_{\hat{\rho}}(k)]}}{M_{\hat{\rho}}(k)} \approx \frac{\sqrt{[(2k)! - k!^2] \zeta_{p+1}(4k)}}{\zeta(2k) + (k!-1)
\zeta_{p+1}(2k)}, \Label{restrictedratio}
\end{equation}
where we defined the partial zeta function as $\zeta_{p+1}(k) = \sum_{m=p+1}^{\infty} m^{-k}$. We may estimate $\zeta_{p+1}(k)$
from below by $\int_{p+1}^{\infty} \frac{dm}{m^k}$ and from above by $\int_p^{\infty} \frac{dm}{m^k}$, from which we find
\begin{equation}
\frac{1}{k-1} \frac{1}{(p+1)^{k-1}} < \zeta_{p+1}(k) <  \frac{1}{k-1} \frac{1}{p^{k-1}}. \Label{zetaestimate}
\end{equation}

For small values of $k$ (\ref{restrictedratio}) does not depend on $N$, and the fluctuations are small with a suitably chosen
$p$. To see this, we estimate
\begin{equation}
\left( \frac{\sqrt{E[M_{\hat{\rho}}(k)]}}{M_{\hat{\rho}}(k)} \right)^2 < [ (2k)! - k!^2 ] \zeta_{p+1}(4k) \lesssim
\frac{(2k)!-k!^2}{(4k-1)p^{4k-1}} < \epsilon_k^2,
\end{equation}
which can always be made smaller than the given measurement precision $\epsilon_k$ with a suitably chosen $p$, without $p$ having
to scale with $N$.

The trouble arises for large values of $k$, i.e. $k \propto N^{\alpha}$, as an observer can optimally measure multipoles up to $k
\sim N^{1/4}$. Using (\ref{zetaestimate}) it can be shown that for the fluctuations to be small, one needs to choose $p \gg k$; a
value so high that almost all the states are projected out of the ensemble, invalidating the statistical treatment of the
system.

We have been unable to find a better method of stabilising the multipoles in the canonical ensemble, as the fluctuations in the
excitation numbers are quite severe. However, one possible resolution to this problem might be that, although naively an observer
is able to measure multipoles up to order $k \sim N^{1/4}$, this might not hold after a more thorough analysis. The reason for
this is that when an observer measures high multipoles, high energies are needed and the backreaction of these should not be
neglected. Also, for low energies it is safe to work within the $\frac{1}{2}$-BPS sector, but for large energies one expects
$g_s$ and $\alpha'$ corrections, which might induce a much stricter limit than $k\sim N^{1/4}$ for the measurable multipoles. If
this was the case, the method of restricting the fluctuations described here could be enough to stabilise the ensemble
sufficiently; a possibility we shall not analyse in more detail in this note.

\section{Discussion}
\label{discussion} In this note we proposed a criterion that a Ramond ground state in the D1--D5 CFT has to satisfy in order to
have a semi-classical gravity dual. This proposal was based on the observation that the data characterizing the CFT state
manifests itself as a set of multipoles in the gravity side. Thus any CFT state having a semi-classical gravity dual has to be
such that the multipoles associated to it do not have a large quantum variance. In particular, we showed that the density matrix
associated to the canonical ensemble is not `sufficiently classical' to admit a semi-classical description, and analysed a possible
way of modifying the ensemble to curb these fluctuations.

Furthermore we showed that while our criterion restricts the states that can have semi-classical duals, certain purely quantum
mechanical aspects can be manifest in the semi-classical gravity dual. An example of this is the observation that quantum
interference in the CFT can give rise to new, measurable, terms in the asymptotic expansion of the metric.

\paragraph{Comparison with LLM:} Since the story proposed in this note closely parallels the one developed in \cite{us,usbig,pettorino} for
the LLM system, it is interesting to analyse the similarities and differences in these systems.

In both cases the relevant states in the CFT's can be described in terms of excitations in a harmonic potential; only in the LLM
case the excitations are fermionic. Thus a basis state is specified by an ordered set of excitation numbers: $\lambda_1 < \ldots
< \lambda_N$, and in \cite{us} we showed that in the expansion of the metric these integers appear in moments $M^{LLM}_k =
\lambda_1^k + \ldots + \lambda_N^k$, which should be compared with the multipoles (\ref{multipoles}) found here. Thus in the LLM
case it is the highest excitations that contribute the most, while in our case the lowest twists are most strongly manifest in
the gravity side. This difference is presumably due to the fractionalization present in the D1--D5 system. In both cases,
however, the CFT data is arranged into a set of moments/multipoles in the gravity side. This analogy extends to superpositions;
in both cases interference terms can be measurable for an asymptotic observer, and some terms in the metric expansion only appear
for states that are superpositions of occupation number eigenstates.

Another similarity between the two systems is apparent in the treatment of the canonical ensemble. In the LLM case it was found
that states with a few highly excited particles, though few in number, contributed disproportionably to the ensemble. Therefore
the ensemble was modified by fixing the highest excitation to be a given number $N_c$ \cite{pettorino}, and the fluctuations in this modified
ensemble were sufficiently constrained to yield the correct stretched horizon for the superstar geometry of \cite{superstar}. In our
case, we found that the fluctuations in the first excitation numbers rendered the ensemble ill-suited for use with the
CFT-to-gravity mapping, and tried to solve this by fixing the first excitation numbers\footnote{The fact that in the LLM case it
was sufficient to fix only one excitation can be traced back to the fact that the excitations are ordered, and thus fixing the
highest will also affect the others.}. Unfortunately, we found that to stabilise the high multipoles, one has to fix so many
excitations that one loses the statistical description of the system.

One final difference between the two systems is that, owing to the fact that in LLM one has a two dimensional phase space and
fermionic excitations, in LLM one can compute the entropy of any spacetime geometry in a very elegant manner. It is not clear if
this can be done in our case, though it would be very interesting if it could be done.

\paragraph{Acknowledgements:} Most of the methods used in this note were developed in \cite{usbig}. I am very grateful to my
collaborators in that paper; Vijay Balasubramanian, Bartek Czech, Don Marolf and Joan Sim\'on, for all their help and support.
I am also thankful to Jan de Boer for helpful comments. I am supported by DOE grant DE-FG02-95ER40893 and a grant from the
Academy of Finland.

\appendix
\section{Combinatorics}
\label{combi}

In this appendix we'll provide a prescription for computing an arbitrary order of the integral in (\ref{basisexp}). Thus we need
to compute
\begin{equation}
\sum_{p=0}^n (-1)^p 2^{n-2p} \binom{2n-p}{p} \! \! \! \! \! \!
\sum_{\stackrel{\stackrel{k_1,\ldots,k_p}{l_1,\ldots,l_p}}{m_1,\ldots,m_{2(n-p)}}} \! \! \! \! \! \! \frac{\delta(\sum_i
(k_i+l_i) + \sum_j m_j)}{\sqrt{|\prod_i k_il_i \prod_j m_j|}} \cdot \mathcal{I}_{\{N_s^a\} }\left[\prod_{i=1}^p \left(
\vec{d}_{k_i}\cdot \vec{d}_{l_i} \right)\prod_{j=1}^{2(n-p)} \left( \vec{e} \cdot \vec{d}_{m_j}\right)\right],
\Label{funcintegral}
\end{equation}
where we have defined the functional integral
\begin{equation}
\mathcal{I}_{\{N_s^a\} }[g(\vec{d})] \equiv \mathcal{N} \int_{\vec{d}} \prod_{s=1}^{\infty} \prod_{a=1}^4 e^{-d_s^ad_{s}^{a*}}
\left( d_s^a d_{s}^{a*} \right)^{N_s^a} g(\vec{d}).
\end{equation}
A basic property of $\mathcal{I}_{\{N_s^a\} }$ is that it factorizes in $s$ and $a$, and we can compute\footnote{Actually, the
computation gives $(N_k^a+1)_r$, but to properly account for the anti-normal ordering prescription we need to translate $d_k^a
d_k^{a*} \to d_k^a d_k^{a*} -1$, after which we get $(N_k^a)_r$.  See \cite{jan} for more details.}
\begin{equation}
\mathcal{I} \left[ (d_k^a d_{-k}^a)^r \right] = \frac{1}{\pi N_k^a!} \int_{d_k^a, d_k^{a*}} e^{-d_k^ad_k^{a*}} \left( d_k^a
d_k^{a*} \right)^{N_k^a+r} = N_k^a (N_k^a+1) \ldots (N_k^a+r-1) = (N_k^a)_r, \Label{basicint}
\end{equation}
where $(x)_n$ is the Pochhammer symbol.

\paragraph{General method:} We see that the integral we need to compute is simply a product of gaussian integrals,
made complicated by the combinatorics of the indices. The integral clearly can be non-zero only when for every index $q$ there is
corresponding index $-q$, i.e. the $2n$ indices $\{ k_i,l_i,m_j\}$ are split into pairs and there are thus only $n$ free indices.
Let us first treat the case where no two pairs share the same value $|q|$. Thus the set of indices is $\{ q_1,\ldots ,
q_N,-q_1,\ldots , -q_N\}$.  The number of times each of these terms appears in the sums over $\{ k_i,l_i,m_j\}$ is $2n!$, since
$k_1$ can be any of the $\pm q_i$, $k_2$ has $2n-1$ options and so on. However, this would completely fix the ordering of the
indices, which we do not want to do; we divide by $n!$, so that $q_1,\ldots ,q_N$ are unordered. Thus, we should always have a
total of  $\frac{(2n)!}{n!}$ terms with all the $q_i$ different.

\begin{figure}
\begin{center}
\includegraphics[scale=0.6]{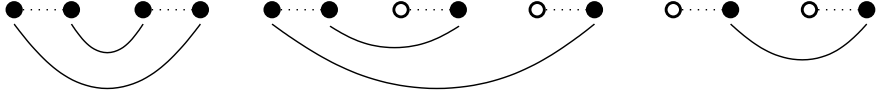}
\end{center}
\caption{\small{Graphical method for writing the argument of the functional integral. Portrayed is the $n=5$, $p=3$ case and one possible pairing.)}}
\label{fig1}
\end{figure}

Next we need to address how the pairings are distributed among the indices $\{k_i,l_i,m_j\}$. All distributions are clearly not
equal, as can be seen from the argument of the functional integral in (\ref{funcintegral}). The clearest way of keeping track of
all possibilities is a graphical representation, and in figure \ref{fig1} we have illustrate the $n=5$, $p=3$ case. In the figure
each solid circle corresponds to a $d$ and each empty circle corresponds to an $e$. The dots between two circles indicate inner
product, i.e. contraction of the $\mathbb{R}^4$ indices.

We need to sum over all possible pairings of indices; we've have drawn one such pairing into the figure, showing with the looping
lines which indices form pairs. We also need to keep track of the $\mathbb{R}^4$ index structure; by following the lines and the
inner products in figure \ref{fig1}, we see that the `strings' created by these lines come in two varieties: `closed' and `open'.
By closed we mean any loop such as the one connecting the left four $d$'s in the figure, while open loops always end in $e$'s
(empty circles) on both ends. Thus there is one closed and two open loops in the figure.

Next we need to see how these loops contribute; this is easiest to do by considering the example in the figure and computing the
contribution from the closed loop and the middle (open) loop. Due to the factorization these can be computed separately, and we
get
\begin{equation}
\left\{ \begin{array}{ll}  \textrm{Closed (left): } & \mathcal{I}[ (\vec{d}_{k_1} \cdot \vec{d}_{l_1}) (\vec{d}_{-l_1} \cdot
\vec{d}_{-k_1}) ] = \sum_{a,b=1}^{4} N_{k_1}^a N_{l_1}^b\delta_{ab} = 4 N_{k_1} N_{l_1}, \\
\textrm{Open (middle): } & \mathcal{I}[ (\vec{d}_{k_3} \cdot \vec{d}_{l_3}) (\vec{e} \cdot \vec{d}_{-l_3}) (\vec{e} \cdot
\vec{d}_{-k_3}) ] = \sum_{a=1}^{4} N_{k_3}^a N_{l_3}^a e_a^2 = N_{k_3} N_{l_3},
\end{array} \right. \Label{closedopen}
\end{equation}
from which we see that closed loops get a factor of $4$ from the index structure, while open ones get $\vec{e}^{\, 2}$, which is
unity. (Note that we are always dealing with the case where the occupation numbers don't depend on direction, i.e. $N_k^a =
N_k$.) Now we are ready to deal with all the cases where no two pairs coincide.

The case where two or more pairs coincide is very similar; the only real difference is the the number of terms we expect. Let us
assume we have $n$ pairs, two of which coincide, i.e. $q_i = q_j$ for some $i$ and $j$. In this case the total number of terms is
$\frac{(2n)!}{(n-2)!2!2!}$, where $\frac{(2n)!}{2!2!}$ is the total number of terms\footnote{For example, when no pairs
coincided, the term $d_{q_i} d_{q_j}$ could come from $k_1=q_i$ and $k_2=q_j$, or $k_1=q_j$ and $k_2=q_i$. Now that $q_i=q_j$
there is only one term, $k_1=k_2=q_i=q_j$; thus we need to divide by the degeneracies.}. We again divide by $(n-2)!$ to make sure
the $q_k$ are unordered. More complicated cases can also be worked out similarly.

\subsection{The $n=2$ case explicitly}
\begin{figure}
\begin{center}
\includegraphics[scale=0.6]{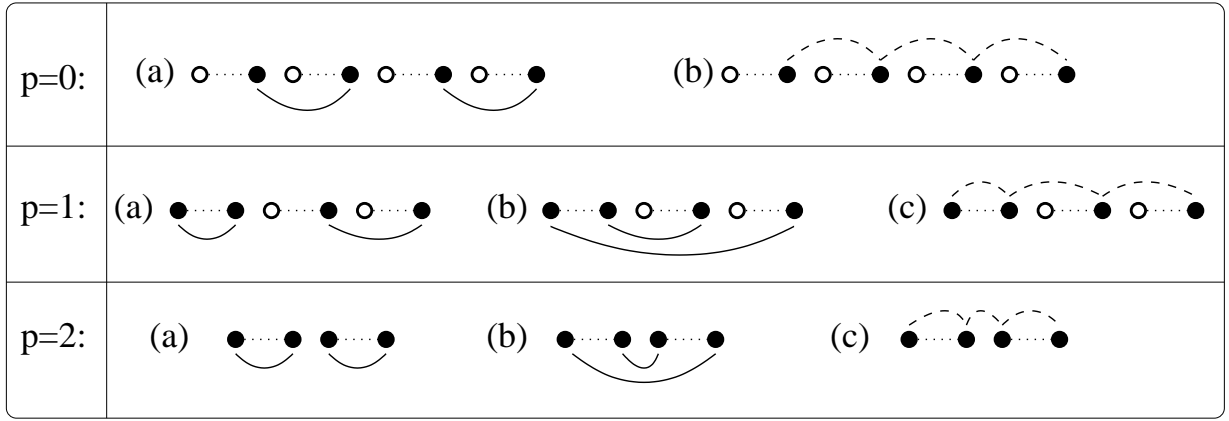}
\end{center}
\caption{\small{All possible `topologically' distinct pairings for $n=2$. The dashes lines above the dots indicate that the dots
connected by these lines share the same index (up to sign). }} \label{fig2}
\end{figure}

To illustrate the method explained above, we shall now work out the case $n=2$ in some detail\footnote{The $n=1$ case, which
turns out to vanish, is too simple and does not illustrate the method particularly well.}. We see from (\ref{funcintegral}) that
we need to compute
\begin{eqnarray}
& & 4 \sum_{m_1,\ldots,m_4} \frac{\delta ( \ldots)}{\sqrt{|m_1m_2m_3m_4|}} \mathcal{I}[  (\vec{e}\cdot \vec{d}_{m_1}
)(\vec{e}\cdot \vec{d}_{m_2} )(\vec{e}\cdot \vec{d}_{m_3} )(\vec{e}\cdot \vec{d}_{m_4} ) ] \nonumber \\ & & - 3
\sum_{k,l,m_1,m_2} \frac{\delta( \ldots)}{\sqrt{|klm_1m_2|}} \mathcal{I}[(\vec{d}_k\cdot\vec{d}_l) (\vec{e}\cdot\vec{d}_{m_1})
(\vec{e} \cdot \vec{d}_{m_2})] + \frac{1}{4} \sum_{k_1,k_2,l_1,l_2} \frac{\delta(\ldots )}{\sqrt{|k_1k_2l_1l_2|}} \mathcal{I}[
(\vec{d}_{k_1}\cdot \vec{d}_{l_1}) (\vec{d}_{k_2} \cdot \vec{d}_{l_2})], \nonumber
\end{eqnarray}
where the terms correspond to $p=0,1,2$ respectively. We'll compute each term separately; all the possible `topologically'
different pairings are drawn in figure \ref{fig2}, and we'll refer to them in the equations as ( \, \, $p=0:\, \,  (a)\, \, )$
etc.

\paragraph{The $p=0$ term:} This is the easiest term and readily gives
\begin{eqnarray}
& & 4 \left\{ (\, \, p=0: \, \, (a)\, \, ) \cdot 3\cdot 2^2 + (\, \, p=0: \, \, (b)\, \, ) \cdot 6 \right\} \nonumber \\ & & =
4\left\{ 12 \sum_{\neq} \frac{1}{m_1m_2} N_{m_1}^aN_{m_2}^b e_a^2 e_b^2 + 6 \sum \frac{1}{m^2} N_m^aN_m^be_a^2e_b^2 \right\}
\nonumber \\ & & = 4\left\{ 12 (\sum \frac{N_m}{m})^2 - 6 \sum \frac{N_m^2}{m^2} \right\} = 48 M_1^2 - 24 M_2.
\end{eqnarray}
This requires some explanation. The sums are over all indices ($\{k_i,l_i,m_j\}$) present and run from 1 to infinity, and
$\sum_{\neq}$ is shorthand for $\sum_{m_1 \neq m_2}$. Sums over the $\mathbb{R}^4$ indices are also present, though we've
suppressed them. The degeneracies on the first line are as follows: 3 is due to $m_1$ being able to pair up with any of the three
other indices, and $2^2$ is due to there being two pairs, in each of which the positive index can be chosen in two ways. In the
second term, $6 = \binom{4}{2}$ is the number of ways two of the four indices can be chosen to be positive. Also note that these
degeneracies coincide with the number of terms as given earlier, namely $12 = \frac{4!}{2!}$ and $6 = \frac{4!}{2!2!}$. Checking
that this is always satisfied is a vital consistency check to make sure the degeneracies are taken into account correctly. We
should also point out that from the formalism above it is clear that the answer can always be given as a sum of the multipoles,
such that the powers are correct, for instance $M_2$ or $M_1^2$ here.

\paragraph{The $p=1$ term:} For the remaining terms, we only give the beginning and the end of the computation; using
(\ref{closedopen}) it is straightforward to fill in the missing steps. The $p=1$ term gives
\begin{eqnarray}
& & -3 \left\{ ( \, \, p=1: \, \, (a)\, \, ) \cdot 4 + (\, \, p=1: (b) \, \, ) \cdot 2\cdot 4 + (\, \, p=1: \, \, (c) \, \,
)\cdot 6 \right\} \nonumber \\
& & = -72 M_1^2 + 18 M_2,
\end{eqnarray}
where we again check that the degeneracies are correct: $4+2\cdot4=12$ and $6$, which is correct.

\paragraph{The $p=2$ term:} Finally, for $p=2$ we get
\begin{eqnarray}
& & \frac{1}{4} \left\{ ( \, \, p=2:\, \, (a) \, \, )\cdot 4 + (\, \, p=2: \, \, (b) \, \, )\cdot 2 \cdot 4 + (\, \, p=2:\, \,
(c) \, \, )\cdot 6 \right\} \nonumber \\
& & = 24 M_1^2 -6 M_2,
\end{eqnarray}
where again the degeneracies match.

Putting these results together we get that the $\frac{1}{r^6}$ term in the asymptotic expansion of the metric is $-12 \frac{Q_5
\mu^4}{r^6}M_2$, as given in (\ref{asympfirst}). Note that the $M_1^2$ terms cancel, leaving only $M_2$.  At the $n=3$ level, one
can show that the $M_1^3$ and $M_1M_2$ terms cancel, leaving only the $M_3$ term. It is tempting to conjecture that this
cancellation always happens, but we've been unable to show this. Nevertheless, the arguments of this paper are not sensitive to
whether terms like $M_1^k$ etc. are present at level $k$ along with the $M_k$ term.

\section{Expansion of $f_1$}
\label{sec:f1exp} For completeness we will also compute the asymptotic form of the $f_1$ function (\ref{quantumf1}). Since $f_1$
differs from $f_5$ only by inclusion of an $|\vec{F}'(s)|^2$ term, we can follow the same recipe as for $f_5$, and we find
\begin{eqnarray}
f_1 & = & 1 + \frac{Q_5}{r^2} \frac{2\pi^2\mu^2}{L^2} \sum_{n=0}^{\infty} \left(\frac{\mu}{r}\right)^{2n} \sum_{p=0}^n (-1)^{p+1}
2^{n-2p} \binom{2n-p}{p} \! \! \! \! \sum_{\stackrel{\stackrel{k_1,\ldots,k_{p+1}}{l_1,\ldots,l_{p+1}}}{m_1,\ldots,m_{2(n-p)}}}
\! \! \! \! \delta(\sum_i (k_i+l_i) + \sum_j m_j) \cdot \nonumber \\ & \cdot & \frac{k_{p+1}l_{p+1}}{\sqrt{|\prod_i k_il_i
\prod_j m_j|}} \mathcal{N} \int_{\vec{d}} \prod_{k=1}^{\infty} \prod_{a=1}^4 e^{-d_k^ad_{k}^{a*}} \left( d_k^a d_{k}^{a*}
\right)^{N_k^a} \prod_{i=1}^{p+1} \left( \vec{d}_{k_i}\cdot \vec{d}_{l_i} \right)\prod_{j=1}^{2(n-p)} \left( \vec{e} \cdot
\vec{d}_{m_j}\right). \Label{f1exp}
\end{eqnarray}
The difference to the expansion of $f_5$ is the inclusion of a factor $-\frac{2\pi^2\mu^2}{L^2} k_{p+1}l_{p+1} \vec{d}_{k_{p+1}}
\cdot \vec{d}_{l_{p+1}} $. Note that using equations (\ref{Qs}), (\ref{mu}) and (\ref{L}), we can write
$Q_5\frac{2\pi^2\mu^2}{L^2} = \frac{Q_1}{2N}$.

\paragraph{The $n=0$ term:} The first term is given by
\begin{equation}
-\frac{Q_1}{r^2} \frac{1}{2N} \sum_{k,l} \delta(k+l) \frac{kl}{\sqrt{|kl|}} \mathcal{I}[(\vec{d}_k\cdot \vec{d}_l)] =
\frac{Q_1}{r^2} \frac{1}{2N} \cdot 2\sum_{k=1}^{\infty} \sum_{a=1}^4 kN_k^a = \frac{Q_1}{r^2},
\end{equation}
which is of course the expected result.

\paragraph{The $n=1$ term:} At the $n=1$ level we have two terms: $p=0,1$. The first one yields
\begin{eqnarray}
-\frac{Q_1\mu^2}{r^4} \frac{1}{2N} 2 \sum_{k,l,m_1,m_2} \delta(k+l+m_1+m_2) \frac{kl}{\sqrt{|klm_1m_2|}}
\mathcal{I}[(\vec{d}_k\cdot \vec{d}_l)(\vec{e}\cdot\vec{d}_{m_1})(\vec{e}\cdot\vec{d}_{m_2})],
\end{eqnarray}
and we see that the possible pairings are just those from the second row of figure (\ref{fig2}). However, the pairing (b) does
not contribute in this case; the reason is that if $k$ and $l$ are independent the sums will yield zero as the summand is odd in
both $k$ and $l$. Thus $k$ and $l$ will always have to be linked to produce a contribution. Thus we get
\begin{eqnarray}
& & -\frac{Q_1\mu^2}{r^4} \frac{1}{N} \left( ( \, \, (a) \, \, ) \cdot 4 + ( \, \, (c) \, \,) \cdot 2 \right)
\nonumber \\
& & = -\frac{Q_1\mu^2}{r^4} \frac{1}{N}  \left( - 4\sum_{k\neq m} \frac{k}{m} N_k^aN_m^be_b^2 - 4 \sum_{k=1}^{\infty} \frac{k}{k}
N_k^aN_k^b e_b^2 + 2 \sum_{k=1}^{\infty}
\frac{k}{k} N_k^aN_k^a e_a^2 \right) \nonumber \\
& & = \frac{Q_1\mu^2}{r^4} \frac{1}{N} \left( 4 \sum_{k,m} \frac{k}{m} N_kN_m - 2 \sum_{k=1}^{\infty} \frac{k}{k} N_k^aN_k^a
e_a^2 \right) \nonumber \\
& & = \frac{Q_1\mu^2}{r^4} \left( 4 M_1 - \frac{2}{N} \sum_{k=1}^{\infty} N_k^2 \right).
\end{eqnarray}
For the $p=1$ term we see that the possible pairings are given on the third line of figure \ref{fig2}, except that (b) again does
not contribute, for the same reason as stated above. The computation proceeds as above and after some algebra we get
\begin{equation}
\frac{Q_1\mu^2}{r^4} \frac{1}{2N} \frac{1}{2} \left( (\, \, (a) \, \,) \cdot 4 + (\, \, (c) \, \,)\cdot 6\right) = \ldots =
\frac{Q_1\mu^2}{r^4} \left( - 4 M_1 + \frac{2}{N} \sum_{k=1}^{\infty} N_k^2 \right).
\end{equation}
Thus we see that the $p=0$ and $p=1$ terms cancel, and at level $n=1$ there is no contribution, which is exactly what happened
for the $f_5$ expansion as well.

\paragraph{The $n=2$ term:} Finally, we can also compute the $n=2$ term. Here the combinatorics are already somewhat complicated,
so we won't present the computation. However, in the end we can write the expansion to order $\frac{1}{r^6}$ as
\begin{equation}
f_1 = 1 + \frac{Q_1}{r^2} - \frac{Q_1\mu^4}{r^6} \left( 12 M_2 - \frac{16}{N} \sum_{k=1}^{\infty} \frac{N_k^3}{k} \right) +
\mathcal{O}(\frac{1}{r^8}).
\end{equation}
Again we see that terms with $M_1$ have cancelled, leaving only $M_2$. However, now we also have a term of the form
$\sum_{k=1}^{\infty} N_k^3/k$, which is not one of the multipoles we have defined. Furthermore, from the formalism we see that
the new objects that can appear are of the form $\sum_k  N_k^n/k^{n-2}$, where the mismatch in powers is due to the factor
$k_{p+1}l_{p+1}$ that came from including $|\vec{F}'(s)|^2$. In principle we should make sure that these quantities don't
fluctuate too much either, but due to their great similarity to the multipoles, it is clear that if we fix the multipoles with
accuracies $\epsilon_k$, then these new objects will also be fixed by some set of frequencies $\epsilon'_k$. Thus we shall not
worry about these objects in this paper.

\end{document}